\documentclass[10pt,conference]{IEEEtran}

\usepackage{xcolor}

\usepackage{multirow} 

\usepackage[pdftex]{graphicx}

\IEEEoverridecommandlockouts
\makeatletter
\def\ps@IEEEtitlepagestyle{
  	\def\@oddfoot{\Footer} 
	}
\let\old@ps@headings\ps@headings
\let\old@ps@IEEEtitlepagestyle\ps@IEEEtitlepagestyle
\def\confheader#1{%
  \def\ps@headings{%
    \old@ps@headings%
    \def\@oddhead{\strut\hfill#1\hfill\strut}%
    \def\@evenhead{\strut\hfill#1\hfill\strut}%
    \def\@oddfoot{\Footer} 
  }%
  \def\ps@IEEEtitlepagestyle{%
    \old@ps@IEEEtitlepagestyle%
    \def\@oddhead{\strut\hfill#1\hfill\strut}%
    \def\@evenhead{\strut\hfill#1\hfill\strut}%
  }%
  \ps@headings%
}
\makeatother
\confheader{} 
\def\Footer{
    {\footnotesize
   \begin{minipage}{\textwidth}
   \centering \scriptsize{DISTRIBUTION STATEMENT A. Approved for public release: distribution is unlimited.\\
   }  \end{minipage}
   }
 }
\begin{document}

\title{Reference-Free Spectral Analysis of EM Side-Channels for Always-on Hardware Trojan Detection\\
}
\author{\IEEEauthorblockN{Mahsa Tahghigh}
\IEEEauthorblockA{Electrical Engineering and Computer Science\\
Howard University,\\
Washington DC, USA,\\
mahsa.tahghigh@bison.howard.edu}
\and
\IEEEauthorblockN{Hassan Salmani, Ph.D.}
\IEEEauthorblockA{Electrical Engineering and Computer Science\\
Howard University,\\
Washington DC, USA,\\
hassan.salmani@howard.edu}
}
\maketitle

\begin{abstract}
Always-on hardware Trojans (HTs) pose a critical risk to trusted microelectronics, yet most side-channel detection methods rely on unavailable golden references. We present a reference-free approach that combines time–frequency EM analysis with Gaussian Mixture Models (GMMs). By applying Short-Time Fourier Transform (STFT) at multiple window sizes, we show that HT-free circuits exhibit fluctuating statistical structure, while always-on HTs leave persistent footprints with fewer, more consistent mixture components. Results on AES-128 demonstrate feasibility without requiring reference models.
\\
\end{abstract}
\renewcommand\IEEEkeywordsname{Keywords}

\section{Introduction}
The security and trustworthiness of microelectronics are foundational to U.S. defense, aerospace, and critical infrastructure systems \cite{Salmani2018}. As integrated circuits become more complex and globally sourced, ensuring that deployed hardware faithfully implements its intended functionality has become increasingly challenging. Hardware Trojans (HTs) – malicious modifications inserted during design, fabrication, or integration – pose a particularly insidious threat. These Trojans can alter functionality, leak sensitive information, degrade reliability, or remain dormant until triggered under specific conditions, undermining mission assurance in safety- and security-critical systems.

Among the various classes of HTs, always-on Trojans represent a uniquely dangerous threat model. Unlike trigger-based Trojans that activate under rare conditions \cite{10.1145/3676536.3689919}, always-on HTs continuously introduce parasitic activity into the circuit, such as additional switching, leakage paths, or covert signal modulation. This persistent behavior increases the likelihood of long-term information leakage or reliability degradation, while also making the Trojan more difficult to isolate using traditional functional testing. Detecting such Trojans is especially challenging in post-deployment or fielded systems, where invasive inspection and exhaustive testing are impractical.

Side-channel analysis, particularly electromagnetic (EM) emission monitoring, has emerged as a powerful non-invasive technique for hardware Trojan detection. EM side-channels provide fine-grained visibility into internal circuit activity and can expose subtle anomalies caused by malicious logic. However, the majority of existing EM-based HT detection approaches rely on golden reference devices, supervised learning, or detailed prior knowledge of the target design. In realistic defense supply chains, such assumptions often do not hold. Trusted reference chips may be unavailable, designs may evolve across fabrication lots, and environmental or operational variability can further obscure direct comparisons, severely limiting the applicability of reference-based techniques. 

He et al. \cite{7994702} proposes a golden-chip-free EM Trojan detection method that compares measured spectra against a simulation-derived reference generated from trusted RTL. While eliminating the need for a physical golden device, the approach still depends on accurate modeling and trusted design visibility, and its evaluation assumes preset Trojan activation, limiting applicability to fielded systems. John et al. \cite{11195049} propose a post-deployment hardware Trojan detector using supervised machine learning on raw power side-channel traces, achieving high accuracy for known Trojans. However, the approach depends on labeled Trojan-infected and Trojan-free data and may not generalize to unseen Trojans or operational drift. Sun et al. \cite{9534884} employ CWT-based EM time–frequency representations combined with transfer learning to classify Trojan-infected and Trojan-free designs with high accuracy. However, the approach relies on labeled training data and fixed time–frequency representations, which may limit robustness to unseen Trojans and operational drift. Prior circuit-level studies have shown that post-fabrication hardware behavior may diverge from design-time assumptions, complicating validation in deployed systems and motivating reference-free analysis methods \cite{cta3980}.

\begin{figure*}[t]
    \centering
    \includegraphics[width=\linewidth]{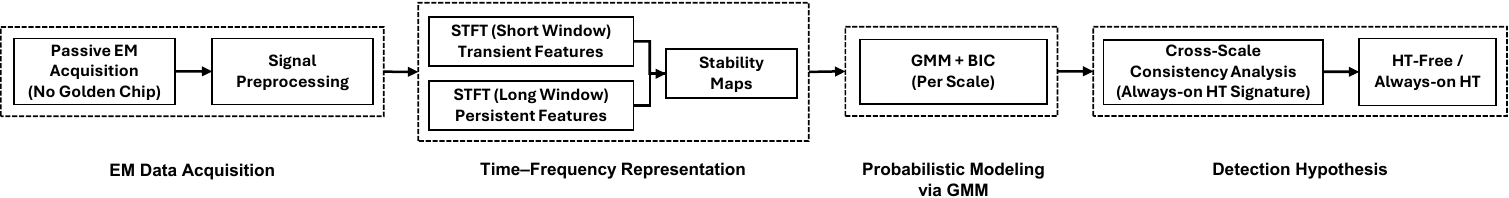}
    \caption{Overview of the proposed reference-free hardware Trojan detection framework, organized into four conceptual steps. (1) Passive EM data acquisition captures side-channel emissions without golden references. (2) Time–frequency representations are generated using multi-resolution STFT to expose transient and persistent spectral behavior. (3) Unsupervised Gaussian Mixture Models (GMMs) with BIC-based model order selection capture statistical structure at each scale. (4)  Detection is performed by evaluating cross-scale consistency, where always-on Trojans exhibit fewer and more stable mixture components than Trojan-free designs.}
    \label{fig:flow}
\end{figure*}

To address these limitations, there is a growing need for reference-free detection frameworks that can operate directly on observed side-channel data without relying on reference models or labeled training sets. Unsupervised statistical learning provides a promising foundation for such approaches, as it enables the characterization of intrinsic structure in side-channel signals without explicit assumptions about normal or malicious behavior. Instead of comparing a device to a known-good reference, a reference-free method evaluates whether the statistical structure of the observed behavior is internally consistent with normal circuit operation. 

In this work, we introduce a reference-free probabilistic framework for detecting always-on hardware Trojans using time–frequency analysis of EM side-channels combined with Gaussian Mixture Modeling (GMM). By applying the Short-Time Fourier Transform (STFT) at multiple window sizes, we examine how the statistical structure of EM emissions evolves across temporal scales. We show that Trojan-free circuits exhibit richer and more variable mixture structures as the time–frequency resolution changes, reflecting diverse operational modes of the circuit. In contrast, always-on Trojans introduce persistent spectral artifacts that suppress this variability, leading to fewer and more consistent mixture components across window sizes. This cross-scale consistency serves as a robust indicator of always-on HT activity, enabling detection without any references, supervised labels, or invasive measurements.

\section{Technical Approach}
We consider a threat model in which an adversary inserts an always-on hardware Trojan into a digital integrated circuit during design, fabrication, or third-party IP integration. Unlike trigger-based Trojans that activate under rare conditions, an always-on Trojan continuously introduces parasitic activity—such as additional switching, leakage paths, or covert modulation—during normal operation. This persistent behavior enables long-term information leakage or system degradation while avoiding detectable functional failures, making always-on Trojans particularly challenging to identify through conventional logic testing or runtime checks. 

We assume a post-deployment detection scenario in which the device under test is already fabricated and operational, and no trusted golden reference chip, labeled Trojan-infected dataset, or accurate simulation-derived EM model is available. The defender has access only to passively measured electromagnetic (EM) side-channel emissions collected during nominal execution of the intended workload. Although Trojan-induced effects may be subtle and masked by noise, process variation, or environmental factors, their always-on nature introduces statistically persistent artifacts across time scales. Accordingly, detection is formulated as identifying abnormal cross-scale statistical consistency in EM emissions rather than matching absolute signatures or performing supervised classification, motivating our reference-free, unsupervised detection framework.

\begin{figure}[b]
    \centering
    \includegraphics{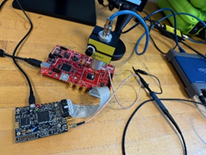}
    \caption{Experimental setup.}
    \label{fig:SetUp}
\end{figure}

\begin{figure*}[t]
    \centering
    \includegraphics[width=\linewidth]{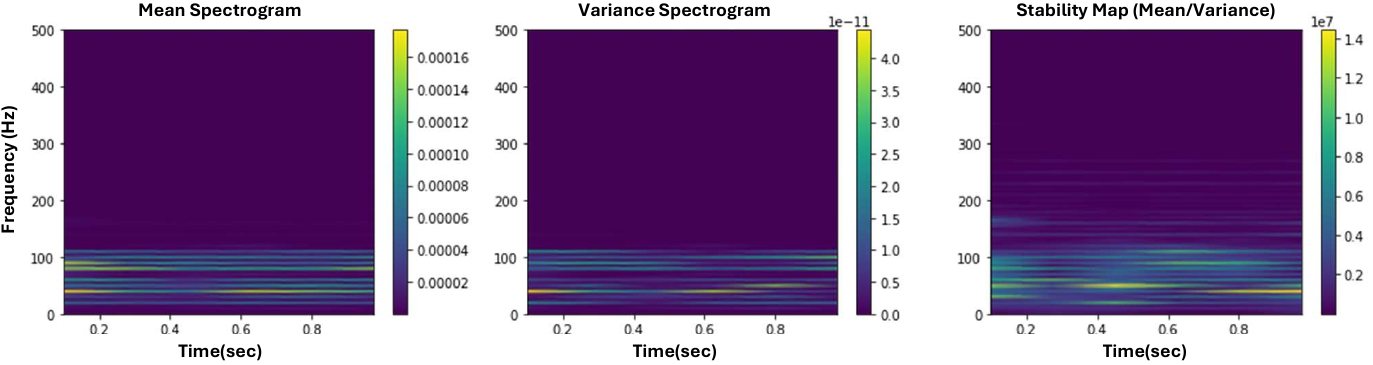}
    \caption{Spectrogram analysis of HT-free AES-128 after applying 500 fixed key and plaintext with the segment length of 200 for STFT.}
    \label{fig:SpectrogramHT-freeAES128-200STFT}
\end{figure*}

\begin{figure*}[t]
    \centering
    \includegraphics[width=\linewidth]{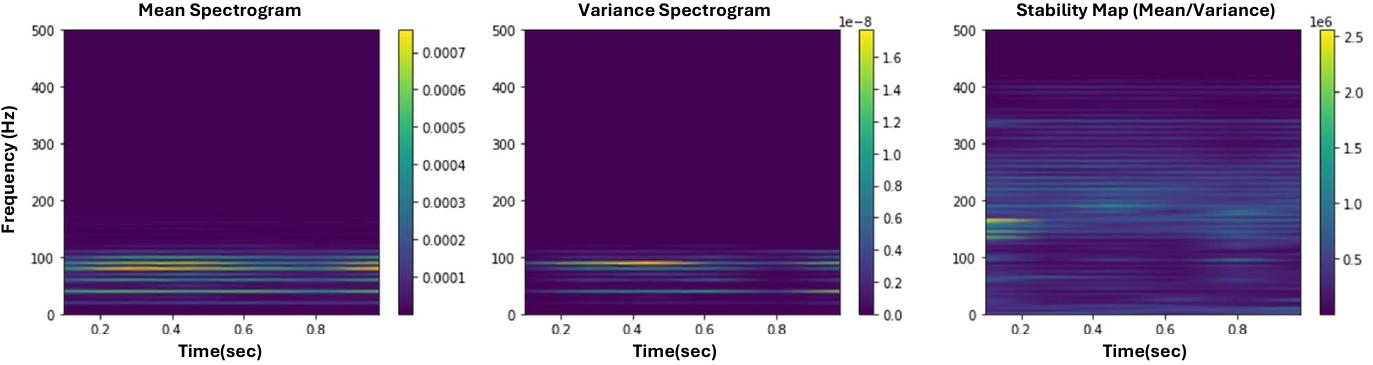}
    \caption{Spectrogram analysis of HT-inserted AES-128 after applying 500 fixed key and plaintext with the segment length of 200 for STFT.}
    \label{fig:SpectrogramHT-InAES128-200STFT}
\end{figure*}

Our methodology combines time–frequency EM analysis with unsupervised probabilistic modeling to detect HTs without reliance on any references, as shown in Figure \ref{fig:flow}. 
\begin{itemize}
    \item \textbf{\textit{EM Data Acquisition}}: We acquire electromagnetic (EM) emanations from a target integrated circuit during normal operation. The process does not assume knowledge of the chip internals or access to a “golden” reference device. Instead, EM signals are collected passively in situ while the device executes its intended functionality, ensuring realistic side-channel conditions where HTs may manifest only sporadically.
    \item \textbf{\textit{Time–Frequency Representation}}: Raw EM traces are transformed into the joint time–frequency domain using the Short-Time Fourier Transform (STFT). To expose both transient and persistent spectral features, we rerun STFT with multiple window sizes. This multi-resolution analysis is critical, as HT activity can leave footprints that appear only at certain temporal scales. The resulting stability maps form a 2D representation of frequency vs. time for each windowing configuration.
    \item \textbf{\textit{Probabilistic Modeling via GMM}}: For each time window, we construct feature vectors of frequency values and their associated stability-map magnitudes. These are analyzed using a Gaussian Mixture Model (GMM) with model order selection via the Bayesian Information Criterion (BIC). The GMM identifies latent statistical structure in the EM emissions, without requiring labels or reference traces.
    \item \textbf{\textit{Detection Hypothesis}}: Our hypothesis is that always-on hardware Trojans leave a persistent spectral footprint in EM emissions that alters the statistical structure of time–frequency representations. By applying Gaussian Mixture Modeling (GMM) across multiple STFT window sizes, we observe a key distinction:
    \begin{itemize}
        \item \textit{HT-free Case}: The number of mixture components fluctuates significantly across window sizes, reflecting that different STFT segmentations capture different operational modes of the circuit (e.g., instruction execution, memory access, interrupts, the rounds of AES execution).
        \item \textit{Always-on HT Case}: The Trojan introduces a stable parasitic signal that suppresses this variability, leading to fewer and more consistent mixture components across window sizes.
    \end{itemize}

Thus, rather than relying on absolute component counts, our method detects anomalies by comparing the consistency of mixture structure across window scales. Persistent, lower-variability GMM profiles are strong indicators of always-on HT activity.

\end{itemize}

\section{Results and Discussion}
To evaluate the proposed reference-free detection framework, we applied it to an AES-128 encryption engine as a representative complex digital workload. The circuit was executed repeatedly under fixed key and plaintext conditions, and 500 EM traces were collected with and without an always-on HT integrated into the design. Figure \ref{fig:SetUp} shows our setup, which uses a Riscure HP EM Probe (1.5 mm) \cite{KeysightInspectorSC4}. EM waveforms are captured using PicoScope software and saved in CSV format.

\begin{figure*}[t]
    \centering
    \includegraphics[width=\linewidth]{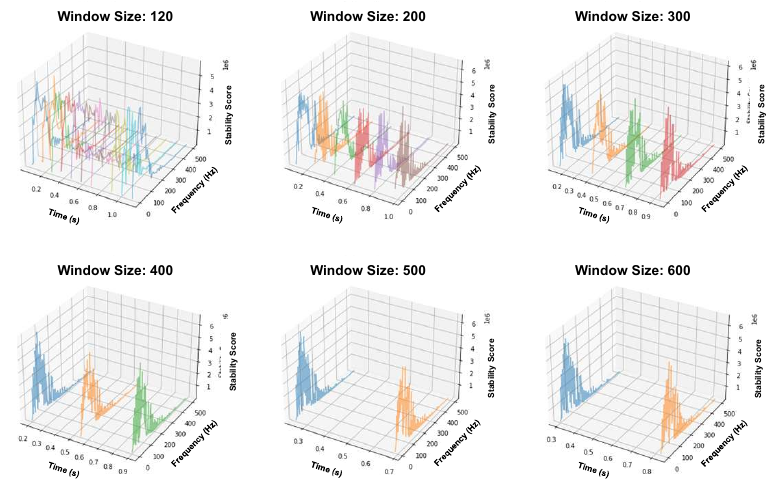}
    \caption{Feature vectors constructed from frequency components and their corresponding stability-map magnitudes for the HT-free AES-128 implementation across multiple STFT window sizes. Each subfigure shows the distribution of stability scores over time and frequency for a given window length.}
    \label{fig:StabilityMapWoHT}
\end{figure*}

Inspired by Code-Division Multiple Access (CDMA), the examined HT is based on AES-T1100 from Trust-Hub. It disperses key-bit leakage across multiple clock cycles using a PRNG seeded with the plaintext to generate a CDMA code, which XOR-modulates the key bits. The modulated sequence is sent to a leakage circuit composed of eight flip-flops, emulating a large capacitance to create a covert power side-channel.  As examples of Step 2 of our proposed methodology, Time– Frequency Representation, Figure \ref{fig:SpectrogramHT-freeAES128-200STFT} and Figure \ref{fig:SpectrogramHT-InAES128-200STFT} show the mean and variance spectrum and stability map (segment length = 200) for STFT, comparing HT-free and HT-inserted AES, respectively. 

\begin{figure*}[t]
    \centering
    \includegraphics[width=\linewidth]{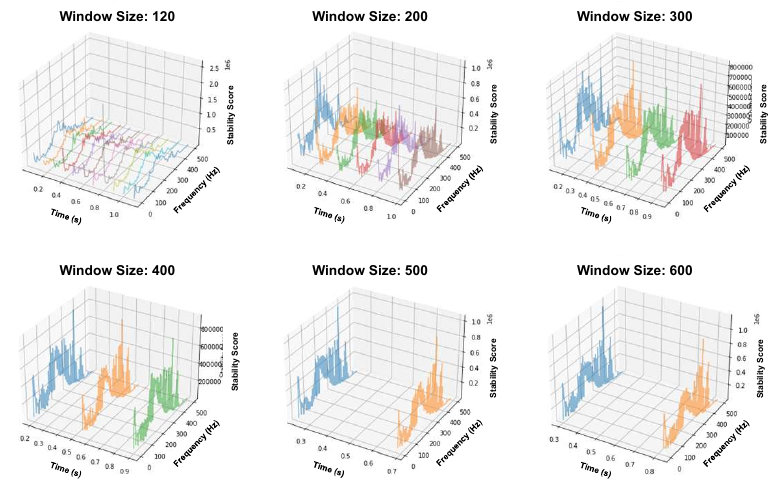}
    \caption{Feature vectors constructed from frequency components and their corresponding stability-map magnitudes for the HT-inserted AES-128 implementation across multiple STFT window sizes.}
    \label{fig:StabilityMapWHT}
\end{figure*}

Continuing with the Step 3, Probabilistic Modeling via GMM, 
Figures \ref{fig:StabilityMapWoHT} and \ref{fig:StabilityMapWHT} present 
the feature vectors of frequency values and their
associated stability-map magnitudes per each time window. 
These are analyzed using a Gaussian Mixture Model (GMM) with model
order selection via the Bayesian Information Criterion (BIC). 
The GMM identifies latent statistical structure in the EM emissions, 
without requiring labels or reference traces. 

Figure \ref{fig:StabilityMapWoHT} illustrates the feature vectors extracted from the HT-free AES-128 implementation, where each feature vector consists of frequency components and their associated stability-map magnitudes for a given STFT window size. As the window length varies, the distribution of stability scores across time and frequency changes noticeably. This scale-dependent behavior reflects the fact that different STFT segmentations emphasize different operational aspects of the circuit, such as instruction execution phases, AES rounds, and memory or control activity. Consequently, the statistical structure captured by the feature vectors varies across window sizes, resulting in heterogeneous patterns that are indicative of normal, Trojan-free operation without persistent parasitic activity.

In contrast, Figure \ref{fig:StabilityMapWHT} shows the corresponding feature vectors for the HT-inserted AES-128 implementation. Across multiple STFT window sizes, the stability distributions exhibit a higher degree of structural consistency, with similar high-stability regions recurring across scales. This persistence suggests that the always-on hardware Trojan introduces a continuous spectral footprint that dominates the time–frequency representation, reducing the variability observed in the HT-free case. As a result, the extracted feature vectors become more uniform across window sizes, leading to fewer and more stable mixture structures when modeled using GMMs. This contrast between scale-dependent variability in Figure \ref{fig:StabilityMapWoHT} and cross-scale consistency in Figure \ref{fig:StabilityMapWHT} directly supports the proposed detection hypothesis that always-on Trojans suppress natural statistical diversity in EM emissions.

\begin{figure}[t]
    \centering
    \includegraphics[width=\linewidth]{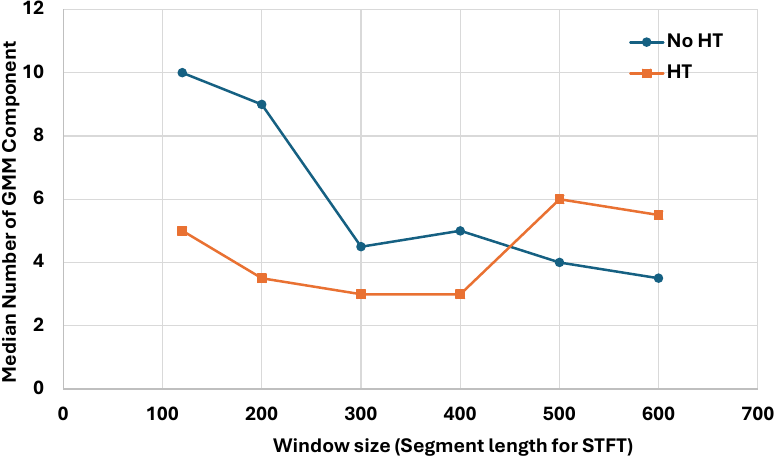}
    \caption{Dependence of the number of GMM-identified components on STFT segment length for AES under HT-inserted and HT-free conditions.}
    \label{fig:DetectionHypothesis}
\end{figure}

Proceeding with the Step 4, Detection Hypothesis, Figure \ref{fig:DetectionHypothesis} 
shows how the number of GMM components varies with 
STFT segment length for HT-inserted and HT-free cases in AES.

To summarize the statistical behavior of the EM emissions at each time–frequency resolution, we compute the median number of GMM components selected by BIC for each STFT window size, shown in Figure \ref{fig:DetectionHypothesis}. The median is used instead of the mean to provide a robust central tendency measure that is less sensitive to outliers caused by noise, transient disturbances, or occasional over-segmentation during mixture modeling. This allows us to capture the dominant statistical structure associated with each window size while minimizing the influence of sporadic fluctuations.

By examining how these median mixture complexities evolve across increasing window sizes, we assess whether the underlying statistical structure of the EM emissions exhibits scale-dependent variability or cross-scale stability. This cross-scale behavior forms the basis of our detection hypothesis, as always-on hardware Trojans are expected to introduce persistent spectral artifacts that suppress natural variability across time–frequency resolutions.

Detection is performed by comparing the relative evolution of mixture complexity across scales rather than by applying a fixed decision threshold on component counts. Across multiple STFT window sizes, the HT-free implementation exhibits a pronounced dependence between time–frequency resolution and statistical complexity. As shown by the median number of GMM components, smaller window sizes (120–200) yield higher mixture complexity, while increasing the window length results in a systematic reduction in the number of components. This behavior indicates that different temporal resolutions capture different operational modes of the circuit, such as instruction execution phases and AES round activity, leading to scale-dependent variation in the latent statistical structure of the EM emissions.

In contrast, the HT-inserted implementation demonstrates suppressed variability across scales. The median number of GMM components remains consistently lower and clustered across window sizes, particularly in the mid-range (200–400), where the mixture complexity collapses to approximately three components. This cross-scale compression suggests that the always-on Trojan introduces a persistent spectral footprint that dominates the time–frequency representation, reducing sensitivity to windowing resolution. The resulting stability of mixture structure across scales provides a strong indicator of always-on Trojan activity and directly supports the proposed detection hypothesis.

\section{Conclusions}
We introduced a reference-free framework for detecting always-on HTs using EM spectrograms and unsupervised GMM analysis. The key finding is that consistent mixture structures across STFT resolutions are strong indicators of Trojan activity. This method avoids any references and invasive inspection, making it well suited for mission assurance in post-deployment and supply-chain-constrained environments where trusted baselines are unavailable.

\centering
\textbf{\\Acknowledgement\\}
This work is supported by the Office of Naval Research under Grant N000142312131.

\bibliographystyle{IEEEtran}
\bibliography{IEEEabrv,GOMACTech_LaTeX}

\end{document}